\documentclass[%
 aip,
 jmp,%
 amsmath,amssymb,
 reprint,%
]{revtex4-1}

\usepackage{graphicx}% Include figure files
\usepackage{dcolumn}% Align table columns on decimal point
\usepackage{bm}% bold math
\usepackage[mathlines]{lineno}% Enable numbering of text and display math
\usepackage{textcomp}
\begin{document}

\preprint{AIP/123-QED}

\title{Reduction of Effective Terahertz Focal Spot Size By Means Of Nested Concentric Parabolic Reflectors}%

\author{V. A. Neumann} 
 \email{v.a.neumann@student.utwente.nl.}
 \affiliation{Department of Physics and Astronomy, The Johns Hopkins University, Baltimore, Maryland, 21218, USA}
 \affiliation{Faculty of Science \& Technology, University of Twente, 7500 AE Enschede, The Netherlands}
\author{N. J. Laurita}%
\author{LiDong Pan}
\author{N. P. Armitage}
 \email{npa@pha.jhu.edu.}
\affiliation{Department of Physics and Astronomy, The Johns Hopkins University, Baltimore, Maryland, 21218, USA}
\date{\today}% It is always \today, today,

\begin{abstract}
An ongoing limitation of terahertz spectroscopy is that the technique is generally limited to the study of relatively large samples of order 4 mm across due to the generally large size of the focal beam spot.  We present a nested concentric parabolic reflector design which can reduce the terahertz focal spot size.  This parabolic reflector design takes advantage of the feature that reflected rays experience a relative time delay which is the same for all paths.  The increase in effective optical path for reflected light is equivalent to the aperture diameter itself.   We have shown that the light throughput of an aperture of 2 mm can be increased by a factor 15 as compared to a regular aperture of the same size at low frequencies. This technique can potentially be used to reduce the focal spot size in terahertz spectroscopy and enable the study of smaller samples.
\end{abstract}
                             
\keywords{Time Domain Terahertz Spectroscopy, Millimeter Wave Spectroscopy, Terahertz Focusing, Parabolic Reflector.}

\maketitle

\section{\label{Introduction}Introduction}

Time-domain terahertz spectroscopy (TDTS) \cite{nuss1998terahertz} has seen tremendous growth is recent years and has found use in many different areas including security applications and biohazard detection \cite{choi2004potential,leahy2007wideband}, detection of protein conformational changes\cite{nagel2002integrated}, structural and medical imaging\cite{hu1995imaging,johnson2001enhanced,mittleman1997t,wang2003t,chan2007imaging}, monitoring of pharmaceutical ingredients \cite{taday2004applications}, and the characterization of materials \cite{kaindl2003ultrafast,heyman1998time,aguilar2012terahertz}. In particular time-domain terahertz spectroscopy has been proven very useful in studying different material systems at low temperatures including superconductors \cite{corson1999vanishing,bilbro2011temporal},  quantum magnets \cite{pan2014low,morris2014hierarchy,laurita2015singlet,bosse2014anomalous}, exciton states in TiO$_2$ nanotubes \cite{richter2010exciton}, and topological insulators \cite{aguilar2012terahertz,aguilar2013aging,wu2013sudden,hancock2011surface}. 

With the photoconductive switches method of TDTS, one typically splits an infrared femtosecond laser pulse along two paths and sequentially excites a pair of photoconductive  ``Auston"-switch antennae. An approximately 1 ps long THz range pulse is emitted by one antenna, focused by mirrors or lenses, transmitted through the sample under test, refocused, and then measured at the other antenna. By varying the path length difference of the two paths, the electric field of the transmitted pulse is measured in the time domain. Ratioing the Fourier transform of the transmission through the sample to that of a reference gives the frequency dependent complex transmission function in a range that typically of order 100 GHz - 3 THz. The complex optical parameters of interest e.g. complex dielectric function or the complex index of refraction can then be obtained from the complex transmission function. However, one of the continuing limiting aspects in using this technique is the typically strong constraints placed on sample configurations.  Measurements are generally performed in transmission, but in order to get adequate signal to noise (particularly at low frequencies) samples must be typically large in the transverse direction.  In most cases sample dimensions must be of order 4 mm across, which is well in excess of the estimates of the diffraction limit of order 0.6 mm.  A principle source of the larger beam spots is from the parabolic aberration that originates from finite size sources and the off-axis parabolic mirrors which are used in most TDTS setups \cite{bruckner2010optimal}.

Because many interesting materials are only found in small crystals, it is important to investigate new ways of measuring small samples.  In this paper we demonstrate a method to compensate for the large beam sizes that come from parabolic aberration and other sources in typical terahertz spectrometers. We use a nested coaxial parabolic reflector arrangement to couple light through a small aperture. The nested coaxial parabolic reflector arrangement is designed to have a one-to-one mapping of light that takes incoming parallel rays, couples them through an aperture as small as 2 mm, and back out into parallel rays (Fig. \ref{Schematic}). It greatly increases throughput and signal to noise by a factor of almost 15 over a simple aperture, making reliable measurements of samples as small as 2 mm possible. The reflector arrangement is compact, compatible with cryogenic arrangements, and can be easily manufactured using standard CNC machining methods.  It may find application in a number of circumstances in TDTS measurements in addition to measuring small samples, including coupling THz radiation into subwavelength waveguides \cite{awad2005transmission}, THz pulse shaping \cite{sato2013terahertz}, and achieving greater overlap between beam sizes in optical-pump THz probe spectroscopies \cite{turner2002carrier}.   Another application of this device could be to significantly reduce the focal spot size in compact TDTS systems consisting of only two off-axis parabolic mirrors. 

\section{\label{Methods}Methods}

\begin{figure}[tbp]
\centering
\includegraphics[width=8.5cm]{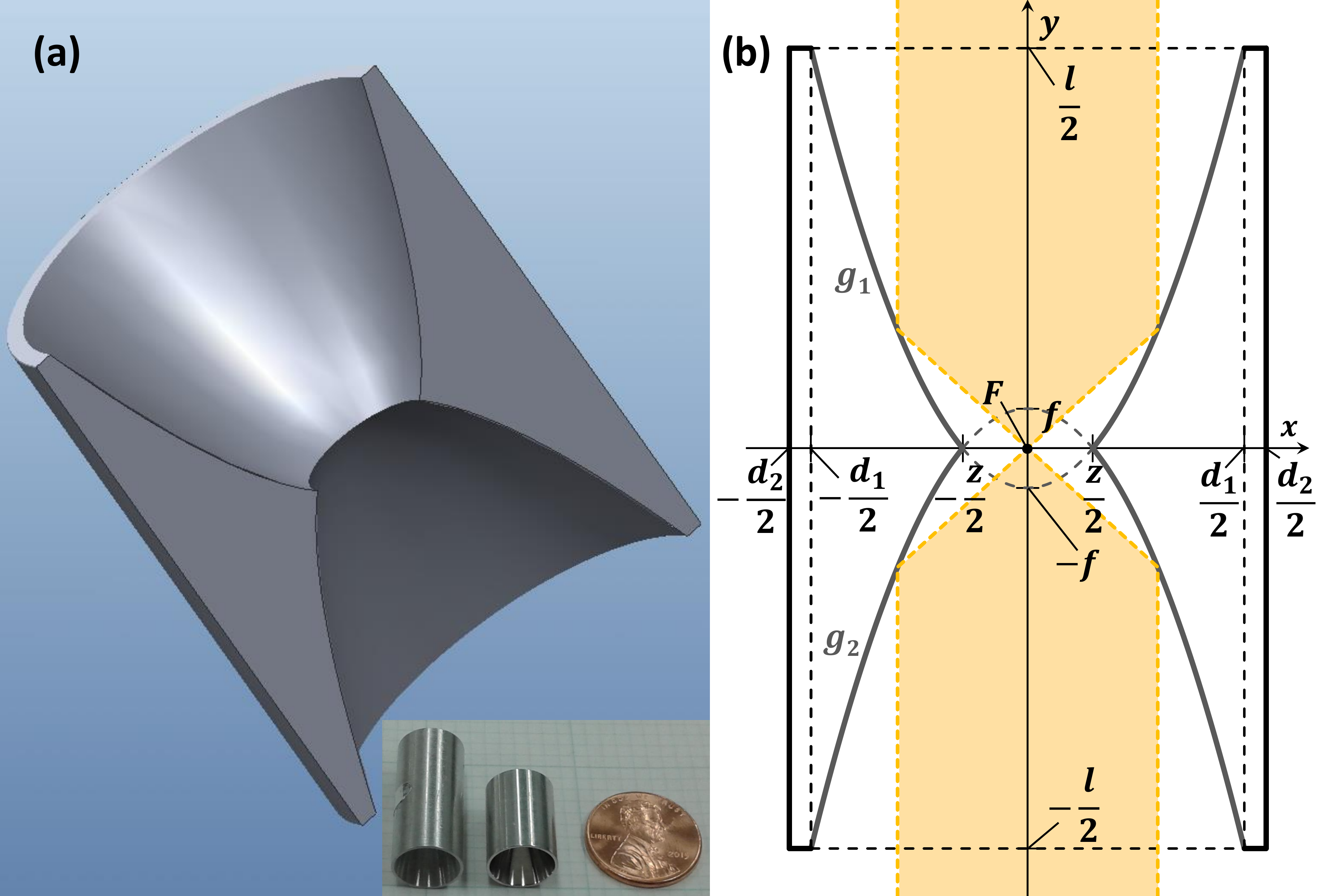}
\caption{(Color online) (a) Three dimensional cross section of the nested concentric parabolic reflector; the inset shows a photo of the 2 mm reflector (left) and 3 mm reflector (right)  (b) Schematic view of the two dimensional cross section of the nested concentric parabolic reflector. The reflective surface is formed by rotating the intersecting parabola $g_1$ and $g_2$ around their axis. $F$ is the focal point of both parabolas, whereas $f$ denotes their focal length. The other dimensions of the parabolic reflector are labeled by the length $l$, the inner diameter $d_1$, the outer diameter $d_2$ and the aperture size $z$. The yellow cone represents a collimated light beam traveling through the device within a ray optics perspective.}
\label{Schematic}
\end{figure}

Fig. \ref{Schematic} shows the design of the nested concentric parabolic reflector. The reflector consists of a cylinder containing two nested paraboloids of revolution, facing in opposite directions, which are aligned in a coaxial manner. The two intersecting paraboloids form a reflecting surface. Fig. \ref{Schematic} (a) and (b) show respectively three and a two dimensional cross sections of the reflector. The paraboloids of revolution are obtained by rotating parabolas $g_1$ and $g_2$ around their axis. The paraboloids $g_1$ and $g_2$ share the focal point $F$, as can be seen in Fig. \ref{Schematic} (b). Within a ray optics perspective, a collimated light beam traveling parallel to the cylindrical surface that enters the top of the reflector, gets reflected by the reflective surface of parabola $g_1$ and consequently becomes focused in the focal point $F$. Subsequently, the light rays hit the reflective surface of parabola $g_2$ and exit the parabolic reflector as a one-to-one mapping of the entering beam. As this mapping can be motivated on the basis of incoming and outgoing waves being time reversed paths of each other, one expects that it is valid within a Gaussian optics perspective as well. Despite the fact that the beam is focused, we expect (in a fashion similar to a Gaussian beam) that the phase fronts of the propagating beam will be flat at the waist e.g in the aperture, allowing general Fresnel equations to be used to determine optical constants of materials from transmission properties.

In the design the parabolas $g_1$ and $g_2$ are given by the formula, 

	\begin{equation}
		y=\pm (\frac{1}{4f}x^2 - f).
		\label{g}
	\end{equation}
	
\noindent In these expressions, $f$ corresponds to the focal length of the parabolas. The dimensions of the reflector can be labeled by several parameters as illustrated in Fig. \ref{Schematic} (b). The values chosen for the diameter at the entrance of the reflector $d_1$ and the effective aperture size in the middle of the reflector $z$ fix the values for the focal length $f$ and the length of the reflector $l$. These parameters are respectively given by the following relations.

	\begin{equation}
		z=4 f
		\label{z}
	\end{equation}
	
	\begin{equation}
		l=2 \bigg[ \frac{d_1^2}{16f}-f \bigg]
		\label{l}
	\end{equation}	

A particular aspect of this design is that -- within a ray optics perspective  -- the relative path length difference of the reflected rays as compared to the rays going straight through the center aperture is a constant independent of the ray trajectory.  Using Eq. \ref{g}, the distance $L$ that a light ray travels after hitting the reflective surface until it reaches the focal point is calculated as 

	\begin{equation}
	L = \sqrt{x^2+y^2}=\sqrt{4fy+4f^2 + y^2} = y+2f
	\label{L}
	\end{equation}

\noindent Here $x$ is the horizontal distance and $y$ is the vertical distance between the point where the light hits the parabolic reflector and the focal point.  It follows that the difference in optical path between two initially parallel rays where one is transmitted directly through the reflector and the other undergoes two reflections is independent of the path through the reflector and given by the expression 

	\begin{equation}
	2 (L - y) = 4f
	\label{diff}
	\end{equation}

\noindent  Hence, the total path length difference between the reflected rays and the light that goes straight through the device is determined by the focal length of parabolas and interestingly is equivalent exactly to the aperture size $z$ itself.

Two nested concentric parabolic reflectors with different dimensions were machined and tested. The outer diameter $d_1$ and inner diameter $d_2$ were equal to 10 mm and 11 mm respectively. The ``2 mm reflector" had an aperture size $z$ of 2.0 mm, which dictates a focal length of $f$ = 0.5 mm and a reflector length of $l$ = 2.40 cm. The ``3 mm reflector" had an aperture $z$ of 3.0 mm and a focal length $f$ and a reflector length $l$ of 0.75 mm and 1.52 cm respectively. The nested concentric parabolic reflectors were machined from 6061 aluminium alloy using standard CNC lathe machining methods to an 0.4 $\mu$m surface roughness. The inset in Fig. \ref{Schematic} (a) shows a photo of the 2 mm reflector (left) and the 3 mm reflector (right). In principle our design could be suitable for even smaller aperture configurations ($<$1 mm), but these could not be produced on the CNC lathe due to machining constraints.  Sinker-type electrical discharge machining (EDM) with electropolishing may be useful in realizing such geometries. The nested concentric parabolic reflectors were tested using a standard TDTS system with Auston switches and off-axis parabolic mirrors (OAP) in an 8f geometry. The parabolic reflector was placed between OAPs 2 and 3. 

\section{\label{ResultsDiscussion}Results and Discussion}

\begin{figure}[tbp]
\centering
\includegraphics[width=8.5cm]{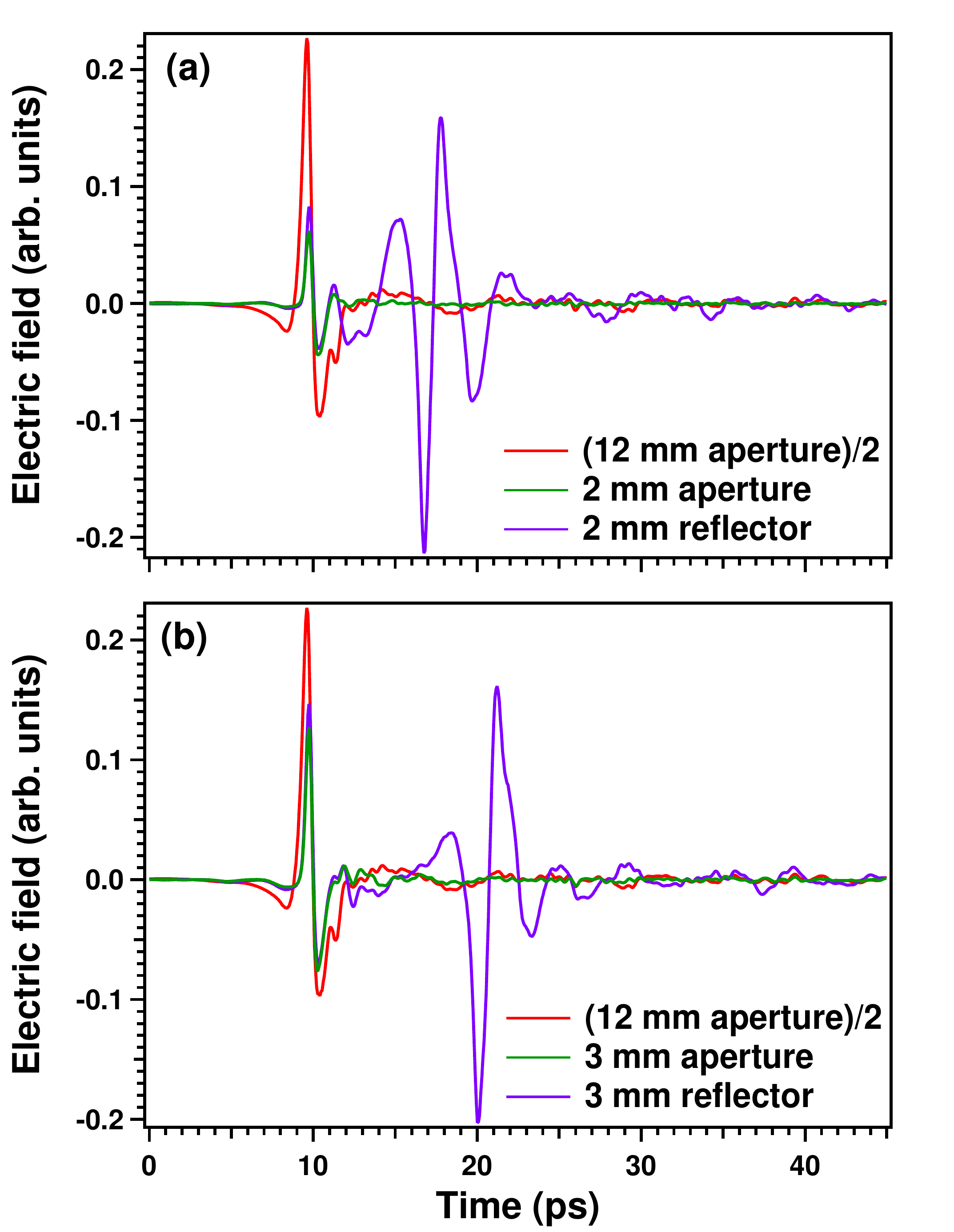}
\caption{(Color online) (a) Comparison of the TDTS time traces of a 12 mm aperture, 2 mm aperture and 2 mm reflector, (b) Comparison of the TDTS time traces of a 12 mm aperture, 3 mm aperture and 3 mm reflector.}
\label{timetraces}
\end{figure}

The electric field as a function of time transmitted through a 12 mm aperture, a 2 mm aperture and the 2 mm reflector are shown in Fig. \ref{timetraces} (a). The 12 mm aperture time trace contains a single peak, as is expected from a typical TDTS time trace. The shape of the 2 mm aperture signal is very similar, expect that the maximum value of the electric field is 7.3 times smaller than for the 12 mm aperture. Thus, the focal spot size of the terahertz beam is significantly larger than 2 mm. The time trace is more complicated for transmission through the reflector. The 2 mm reflector transmission shows a first peak at the same location in time as the 2 mm aperture that is extremely similar to the 2 mm aperture's shape, except that the maximum electric field of the 2 mm reflector peak is slightly higher by a factor of 1.3. This effect is probably caused by small differences in the size of the aperture at the center of the reflector. The initial feature is presumably due to the straight transmission through the aperture with no reflection off the mirror surfaces.  Most striking in the reflector spectrum however is the second feature that shows up slightly later in time. The shape of this second feature is somewhat similar to the shape of the first peak, but the second peak is broader in time, larger in magnitude and phase shifted as compared to the first peak. The time difference between the first and the second peaks is 7.0 ps, which corresponds to a path length increase of 2.1 mm. This path length difference is approximately equal to the aperture size $z$ as predicted in Eq. \ref{diff} above for the reflected peak. It can be concluded that the second peak is due to radiation being reflected by the parabolic surface and subsequently being passed through the aperture at the center of the reflector. The maximum electric field of the second peak, which will be called the reflector peak hereafter, is 2.6 times as large as the first peak. 

Fig. \ref{timetraces} (b) shows the electric field of the transmission through respectively a 12 mm aperture, a 3 mm aperture and the 3 mm reflector as a function of time. Similar trends as for the 2 mm reflector can be observed and in this case the time difference between the first peak and the reflector peak is larger and corresponds to 3.08 mm. Again, this is approximately equal to the size of the aperture $z$ as has been predicted above. The maximum electric field of the second peak is now 1.4 times larger than the first peak. The smaller ratio between these two peaks in the case of the 3 mm reflector indicates that this reflector is less effective at coupling additional light through the aperture in the center. This is presumably expected because a larger percentage of the signal can make it through on the first pass with no reflection.  We note that additional polishing of the 3 mm reflector to a mirror finish did not increase the throughput, but instead induced an additional and erratic phase shift in the signal. This is probably caused by deviations in the parabolic shape which were induced in the polishing process.

\begin{figure}[tbp]
\centering
\includegraphics[width=8.5cm]{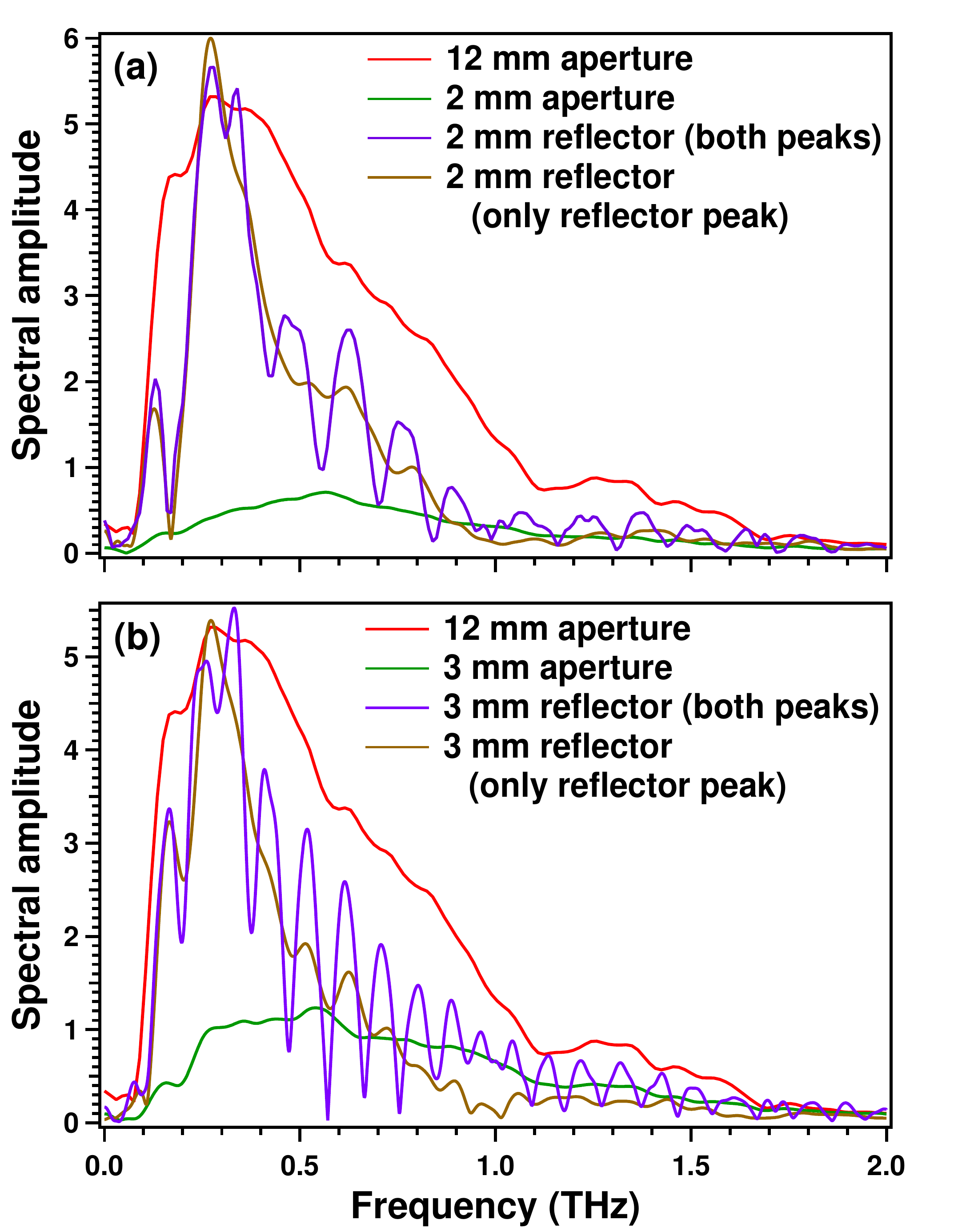}
\caption{(Color online) (a) Comparison of the spectral amplitudes of the throughput of a 12 mm aperture, 2 mm reflector and 2 mm parabolic reflector, (b) comparison of the spectral amplitudes of the throughput of a 12 mm aperture, 3 mm reflector and 3 mm parabolic reflector. The reflector spectral amplitude is analyzed by taking Fourier transform of the whole signal (``both peaks") or only of the second peak (``only reflector peak").}
\label{FFTs}
\end{figure} 

Fig. \ref{FFTs} shows a comparison of the spectral amplitudes, found by taking the Fourier Transform of the time traces in Fig. \ref{timetraces},  of the 12 mm aperture, 2 mm aperture, and 2 and 3 mm reflectors. It can be seen that the 12 mm aperture signal has a much higher spectral amplitude than the 2 and 3 mm aperture. This difference in magnitude is most pronounced at low frequencies ($\sim$0.1 up to $\sim$0.8 THz), because of the large focal spot size of the low frequency light. The focal spot size of the high frequency light is smaller, therefore the difference in spectral magnitude between the apertures is smaller. The reflector signals were analyzed in two different ways. The spectral amplitude obtained by taking the Fourier transform of the complete reflector signal shows strong oscillations. These oscillations are caused by interference between the first peak and the reflector peak in the time trace. These oscillations can be removed by only analyzing the reflector peak. This has been achieved by subtracting the rescaled aperture time trace from the reflector time trace before taking the Fourier transform. As can be seen in Fig. \ref{FFTs}, this removes most of the oscillations. The oscillations cannot be completely removed in this fashion because the shape of the first peak in the reflector and aperture time trace are not identical. 

The maximum spectral amplitude in the case of the reflector is 15 times the amplitude of a regular aperture of the same size in the case of the 2 mm reflector and 5 times for the 3 mm reflector at 0.27 THz. Between 0.20 and 0.35 THz the spectral amplitude of the reflector is of the same order as the the 12 mm aperture. The reflectors significantly increase the signal up to a frequency of 0.7 THz. The reflector peak is broader in time than the first peak as can be seen in Fig. \ref{timetraces}. A peak which is broader in time is composed of lower frequencies and this is indeed the case as can be seen in the spectrum.   Strictly speaking this low frequency radiation does not obey the traditional laws of ray optics, but instead forms a Gaussian beam waist with a large spot size when entering the reflector. The reflectors couple this low frequency light through the hole very efficiently. Light with a frequency above $\sim$1 THz has a much smaller spot size. Most light in this domain makes it through the aperture without reflecting. Therefore, the reflector does not significantly increase the spectral amplitude in this region.  

\section{\label{Conclusion}Conclusion}

We have demonstrated an effective way of increasing the light throughput of a small aperture. It can be concluded that the novel nested concentric parabolic reflector significantly reduces the effective focal spot size of low frequency terahertz radiation by coupling it through a small aperture. The results suggest that a reflector with a smaller aperture than demonstrated in the paper might be even more efficient at increasing the throughput of light. 

This technique could enable the study of smaller samples than has been possible so far by means of TDTS experiments, by placing a sample in a slot machined in the middle of the parabolic reflector. In order to maintain the one-to-one mapping of in-going to out-going rays and the proper focussing at least approximately, the design would presumably have to account for the additional optical path length in the sample under test.  Moreover in order to be able to analyze the data, it is important that the reflector peak does not overlap with the reflections from Fabry-P\'{e}rot interference from multiple reflections in a substrate. Thus, optically thick samples and/or substrates should be used to clearly separate these peaks. 

\section{\label{Acknowledgements}Acknowledgements}

The THz instrumentation development was funded by the Gordon and Betty Moore Foundation through Grant GBMF2628 to NPA. We would like to thank Robert Barkhouser of the Instrument Design Group of The Johns Hopkins University for help with design specifications for the reflectors and aid in Fig. \ref{Schematic} (a).

\bibliographystyle{unsrt}
\bibliography{mybib}{}

\end{document}